\documentclass[prl,preprintnumbers,twocolumn,showpacs,nobalancelastpage]{revtex4}

\usepackage{amsmath,amssymb,amsfonts}
\usepackage{bm,graphicx,color}

\newcommand{\CC}{\mathcal{C}}
\newcommand{\TC}{\text{T}_{\CC}}
\newcommand{\TR}{\text{Tr}}
\newcommand{\BK}{{\bm k}}
\newcommand{\myast}{{{}\hspace*{-0.1em}\ast\hspace*{-0.1em}{}}}
\newcommand{\mydagger}{{\dagger}}
\newcommand{\phdagger}{{\phantom{\mydagger}\!}}

\newcommand{\Gwnull}{Q}\newcommand{\Gweins}{R}
\newcommand{\gwnull}{q}\newcommand{\gweins}{r}
\newcommand{\KERN}{M}
\newcommand{\SUBKERN}{K}
\newcommand{\NF}{\bm{n}_{\!f}}
\newcommand{\bra}[1]{\langle{#1}|}
\newcommand{\ket}[1]{|{#1}\rangle}
\newcommand{\braket}[2]{\langle{#1}|{#2}\rangle}
\newcommand{\expval}[1]{\langle{#1}\rangle}

\begin{document}

  \title{Nonthermal steady states after an interaction quench
    in the Falicov-Kimball model}

  \author{Martin Eckstein}
  
  \author{Marcus Kollar}

  \affiliation{Theoretical Physics III, Center for Electronic
    Correlations and Magnetism, Institute for Physics, University of
    Augsburg, 86135 Augsburg, Germany}
  
  \date{July 18, 2007}

  \begin{abstract}
    We present the exact solution of the Falicov-Kimball model after a
    sudden change of its interaction parameter using non-equilibrium
    dynamical mean-field theory.  For different interaction quenches
    between the homogeneous metallic and insulating phases the system
    relaxes to a non-thermal steady state on time scales on the order
    of $\hbar$/bandwidth, showing collapse and revival with an
    approximate period of $h$/interaction if the interaction is large.
    We discuss the reasons for this behavior and provide a statistical
    description of the final steady state by means of generalized
    Gibbs ensembles.%
  \end{abstract}

  \pacs{03.75.Ss, 05.30.Fk, 71.27.+a%
    \vspace*{-3mm}%
  }

  \maketitle

  How does an isolated quantum-mechanical many-body system develop
  after it is suddenly forced out of thermal equilibrium? Under which
  conditions does it relax to a new steady state, and how fast? Is it
  ergodic so that it reaches a new thermodynamic equilibrium, or does
  the final state depend on the initial state? Recently it has become
  feasible to study these fundamental questions experimentally and
  theoretically. In experiments with ultracold atomic gases
  \cite{Bloch07} it is possible to subject a prepared initial state to
  a rapid change of system parameters. Long observation times are
  possible due to the excellent isolation from the environment. For
  example, Bose-Einstein condensates (BECs) were quenched across the
  superfluid-insulator transition and back \cite{Greiner02a}, their
  collapse and revival after a quench was observed \cite{Greiner02b},
  a quenched spinor BEC was found to exhibit spontaneous symmetry
  breaking \cite{Sadler06}, and a quantum version of Newton's cradle
  was found not to thermalize \cite{Kinoshita06}.

  One might expect that a quenched system with many interacting
  degrees of freedom will relax to a new thermal state, characterized
  only by a few thermodynamic variables such as internal energy and
  particle number. However this may not be the case if the system is
  integrable, because then the final state is constrained by
  infinitely many constants of motion.  Indeed, theoretical studies
  for one-dimensional hard-core bosons \cite{Rigol07,Rigol06}
  (experimentally realized in Ref.~\onlinecite{Kinoshita06}) and for
  the Luttinger model \cite{Cazalilla06} found that these integrable
  systems relax to non-thermal steady states.  Nevertheless for both
  models the final state is described by a generalized Gibbs ensemble
  \cite{Rigol07}, which maximizes the entropy subject to all
  constraints.  On the other hand, for non-integrable and
  unconstrained systems the usual Gibbs ensemble should describe the
  final steady state.  In contrast to this expectation recent
  numerical studies for finite one-dimensional systems of soft-core
  bosons \cite{Kollath07} and spinless fermions \cite{Manmana07} did
  not find thermalization.  While the reasons for this behavior are
  not yet understood, hard-core bosons in two dimensions do thermalize
  as expected \cite{Rigol07b}. Clearly finite-size effects must be
  well-controlled in all such calculations in order to obtain the
  correct behavior at large times.
  
  Dynamical mean-field theory (DMFT) \cite{Metzner89,Georges96}, which
  has become a standard technique for correlated systems in
  equilibrium, can also provide insight into their quantum dynamics,
  e.g., in the presence of time-dependent external fields
  \cite{Schmidt02,Turkowski05+Freericks06}.  DMFT has the conceptual
  advantages that it is formulated in the thermodynamic limit so that
  finite-size lattice effects are eliminated, and that it becomes
  exact for high-dimensional lattices.  As such, it is complementary
  to numerical methods for finite low-dimensional systems.  The
  characteristic features of DMFT for fermions \cite{Georges96} or
  bosons \cite{Byczuk07}, namely a local self-energy derived from a
  local action with self-consistency condition, persist also for
  non-equilibrium situations.
  
  In this paper we use DMFT to study quenches in the interaction
  parameter of the Falicov-Kimball (FK) model. This lattice model
  describes itinerant $c$ electrons and immobile $f$ electrons that
  interact via a repulsive local interaction $U$ \cite{Falicov69}. The
  Hamiltonian is given by
  \begin{align}%
    H
    &=
    \sum_{ij} V_{ij}
    c_i^\mydagger
    c_j^\phdagger
    +
    E_f
    \sum_i
    f_i^\mydagger
    f_i^\phdagger
    +
    U
    \sum_i
    f_i^\mydagger
    f_i^\phdagger
    c_i^\mydagger
    c_i^\phdagger
    \,,\label{eq:FK-model}%
  \end{align}%
  i.e., it is similar to the Hubbard model except that only one
  electron species can hop between lattice sites.  In DMFT the
  effective local action for the $c$ particles is quadratic, so that
  their Green function can be obtained exactly
  \cite{Brandt89,vanDongen90+92}. The equilibrium solution describes
  correlation-induced transitions between metallic, insulating, and
  charge-ordered phases~\cite{Freericks03}.  The FK model proved very
  useful as a guide for the application of DMFT to the Hubbard model.
  It currently plays a similar role for nonequilibrium DMFT, in
  particular since no appropriate real-time impurity solver is yet
  available for the Hubbard model, although, e.g., time-dependent
  numerical-renormalization group \cite{Anders05} is a promising
  candidate.  So far, however, even the self-consistency equation has
  required tremendous numerical effort for a general nonequilibrium
  situation due to lack of time-translational
  invariance~\cite{Turkowski05+Freericks06}. For the investigation of
  an interaction quench we consider a semi-elliptical density of
  states, which leads to a dramatic simplification of the
  self-consistency equation both for the FK and Hubbard model.
   
  We assume that the system is prepared in thermal equilibrium at
  temperature $T$ for times $t$ $<$ $0$; at $t$ $=$ $0$ the
  interaction is suddenly switched from the value $U_-$ to a new value
  $U_+$, so that the time evolution for $t\geq0$ is governed by the
  new Hamiltonian \cite{footnote1}.
  Below we obtain the exact non-equilibrium Green
  function for arbitrary quenches and arbitrary large times.


  {\em Non-equilibrium DMFT.---} %
  The theory is formulated in terms of contour-ordered real-time Green
  functions. In general, this formalism is appropriate to describe an
  isolated system, where the initial state is a density matrix
  \cite{Haug96}.  We use the Keldysh Green functions $G_{ij}(t,t')$
  $=$ $ -i\langle \TC c_i^\phdagger(t)c_j^\mydagger(t')\rangle$, which
  are defined on the contour $\CC$ that runs from a negative
  $t_{\text{min}}$ to a positive $t_{\text{max}}$, then from
  $t_{\text{max}}$ to $t_{\text{min}}$, and finally to
  $t_{\text{min}}-i\beta$ \cite{Turkowski05+Freericks06}. Here
  $\langle\cdot\rangle$ $=$ $\TR [e^{-(H(t_{min})-\mu N)/T}\cdot]$ is
  the thermal expectation value with chemical potential $\mu$ and
    total particle number $N$.  For the FK model the local Green
  function $G(t,t')$ in the homogeneous phase is calculated from a
  local action~\cite{Brandt89,Turkowski05+Freericks06},
  \begin{subequations}\label{eq:local-g}%
    \begin{align}%
      G(t,t') &= -i\frac{ \TR_{c,f} [e^{-\beta H_0} \TC S_1 S_2
        c(t)c^\mydagger(t')] }{ \TR_{c,f} [e^{-\beta H_0} \TC S_1
        S_2]},
      \\
       S_1&= \exp\left(-i\int_\CC\!d\bar{t} \int_\CC\!d\bar{t}' \,
         c^\mydagger(\bar{t}) \Lambda(\bar{t},\bar{t}')
         c(\bar{t}')\right)
       ,\\
       S_2&= \exp\left(-i\int_\CC\!d\bar{t} \, U(\bar{t})
         c^\mydagger(\bar{t})c(\bar{t})f^\mydagger(\bar{t})f(\bar{t})\right)
       ,
    \end{align}%
  \end{subequations}%
  where the operators are in the interaction representation with
  respect to $H_0$ $=$ $(E_f-\mu)f^\mydagger f-\mu c^\mydagger c$,
  and $\hbar$ $=$ $1$.  After
  tracing out the $f$ electrons and setting $w_1$ $=$ $\langle
  f^\mydagger f\rangle$ $=$ $1-w_0$ one has
  \begin{subequations}\label{eq:g-from-lambda}%
    \begin{align}%
      G(t,t')
      &=
      w_0 \Gwnull(t,t')
      +
      w_1 \Gweins(t,t')
      \,,\label{eq:g=b+d}
    \end{align}%
    where $\Gwnull(t,t')$ and $\Gweins(t,t')$ are given by
    (\ref{eq:local-g}) but without $\TR_{f}$ and with
    $f^\mydagger(\bar{t})f(\bar{t})$ replaced by 0 and 1,
    respectively.  From (\ref{eq:local-g}) follow the equations of
    motion
    \begin{align}%
      {}[i\partial_{t}^\CC\! + \mu]
      \Gwnull(t,t')
      - 
      (\Lambda \myast \Gwnull)(t,t')
      &=
      \delta^\CC\!(t,t')
      \,,\label{eq:eqm-d}
      \\
      {}[i\partial_{t}^\CC\! + \mu - U(t)]
      \Gweins(t,t')
      - 
      (\Lambda \myast \Gweins)(t,t')
      &=
      \delta^\CC\!(t,t')
      \,,\label{eq:eqm-b}
    \end{align}%
  \end{subequations}%
  where $(f \myast g)(t,t')$ $=$ $\int_\CC
  d\bar{t}f(t,\bar{t})g(\bar{t},t')$ denotes the convolution,
  $\partial_t^\CC$ the derivative, and $\delta^\CC\!(t,t')$ the delta
  function along the contour~\cite{Turkowski05+Freericks06},
  and the Green functions obey antiperiodic boundary conditions.
  
  In DMFT the contour self-energy is local and its skeleton expansion
  in terms of the contour Green function is the same as that of the
  self-energy of the local problem (\ref{eq:local-g}), determined from
  its Dyson equation
  \begin{align}
    (i\partial_t^\CC+\mu)
    G(t,t')
    -
    ([\Lambda + \Sigma] \myast G)(t,t')
    &=
    \delta^\CC\!(t,t')
    \,.
    \label{eq:local-dyson}
  \end{align}
  On the other hand, the lattice Dyson equation provides a relation
  between the self-energy and the lattice contour Green function
  $G_{ij}(t,t')$,%
  \begin{subequations}\label{eq:lattice-dyson}%
    \begin{align}%
      \!
      (i\partial_t^{\CC}+\mu-\epsilon_\BK)
      G_\BK(t,t')
      -
      (\Sigma \myast G_\BK)(t,t')
      &=
      \delta^{\CC}(t,t')
      \,,\label{eq:lattice-dyson-sigma}%
    \end{align}%
    where $\epsilon_\BK$ are the eigenvalues of the matrix $V_{ij}$.
    In the corresponding eigenbasis the lattice contour Green function
    $G_\BK(t,t')\equiv G_{\epsilon_\BK}(t,t')$ is diagonal and depends
    on $\BK$ only through $\epsilon_\BK$. The self-consistency
    equation
    \begin{align}%
      G(t,t')
      &=
      \int\!d\BK\,G_\BK(t,t')
      =
      \int\!d\epsilon\,\rho(\epsilon)G_\epsilon(t,t')
      \,,\label{eq:lattice-dyson-gloc}%
    \end{align}%
  \end{subequations}%
  then closes the problem, i.e., there are three equations
  (\ref{eq:g-from-lambda}), (\ref{eq:local-dyson}),
  (\ref{eq:lattice-dyson}) for three unknowns $G(t,t')$,
  $\Lambda(t,t')$, $\Sigma(t,t')$. For a general density of states
  $\rho(\epsilon)$ the numerical evaluation of
  (\ref{eq:lattice-dyson}) is expensive, because the integral equation
  (\ref{eq:lattice-dyson-sigma}) must be solved for every integration
  point in (\ref{eq:lattice-dyson-gloc})
  \cite{Turkowski05+Freericks06}.  This problem simplifies
  dramatically for a semi-elliptic density of states $\rho(\epsilon)$
  $=$ $\sqrt{4 V^2-\epsilon^2}/2\pi V$.
  In this case, the Hilbert transform $g(z)=\int\!d\epsilon\,\rho(\epsilon)
  /(z-\epsilon)$ satisfies the equation $zg=1+V^2g$, and this 
  also holds for linear operators~\cite{Eckstein_unpublished},
  e.g., $z$ $=$ $(i\partial_t^{\CC}+\mu-\Sigma)$ and $g$ $=$ $G(t,t')$.
  Thus (\ref{eq:lattice-dyson}) reduces to
  \begin{align*}%
    \!
    (i\partial_t^{\CC}\!+\mu  )
    G(t,t')
    -
    ([\Sigma + V^2G] \myast G)(t,t')
    =
    \delta^{\CC}\!(t,t')
    \,,
  \end{align*}%
  so that, by comparison with (\ref{eq:local-dyson}),
  \begin{align}
    \Lambda(t,t')=V^2G(t,t')
    \,.\label{eq:lambda-bethe}
  \end{align}


  {\em Analytic solution.---} %
  We now solve (\ref{eq:g-from-lambda}) and (\ref {eq:lambda-bethe}) for
  an interaction quench at $t$ $=$ $0$. Because the Hamiltonian does not 
  change for times $t$ $<$ $0$, the Green functions take their
  equilibrium values when both $t$ $<$ $0$ and $t'$ $<$ $0$. 
  We take this as an initial condition in Eq.~(\ref{eq:g-from-lambda})
  and remove the vertical part of the contour by letting 
  $t_{\text{min}}$ $\to$ $-\infty$; correlations
  such as $G(t,t_{\text{min}}-i\tau)$ between times $t$ on the
  real part of the contour and $t_{\text{min}}-i\tau$ on the imaginary part
  vanish in this limit.
  Using Langreth rules \cite{Haug96} 
  we then recast (\ref{eq:g-from-lambda}) into a set of coupled 
  integro-differential equations for the lesser component $G^<(t,t')$ $=$ 
  $i\langle c^\mydagger(t')c(t)\rangle$ and the retarded component 
  $G^R(t,t')$ $=$ $-i\Theta(t-t')\langle \{c^\mydagger(t'),c(t)\}\rangle$.
  Directly from these rules and the fact that any retarded function
  $f(t,t')$ must vanish for $t<t'$, one can see that within these
  equations the retarded Green functions with $t>t'>0$ and $0>t>t'$
  are decoupled from all other components. Moreover, the corresponding
  two sets of equations differ only in the value of $U$, and both
  are translational invariant in time. Thus they can be written in terms
  of the Fourier transforms $g^R_\pm(z)$ ($\pm$ for $t,t'$ $\gtrless$ 
  $0$, respectively) with respect to $t-t'$,%
  \begin{subequations}\label{eq:cubic}%
    \begin{align}%
      g^R_\pm(z) &= w_0\gwnull^R_\pm(z) + w_1\gweins^R_\pm(z)
      \,,
      \\
      \gwnull^R_\pm(z) &= [z+\mu - V^2 g^R_\pm(z)]^{-1}
      \,
      \\
      \gweins^R_\pm(z) &= [z+\mu - V^2 g^R_\pm(z) - U_\pm ]^{-1}
      \,.
    \end{align}%
  \end{subequations}%
  The same set of cubic equations determines the equilibrium Green
  function \cite{vanDongen90+92}, but in the present case $\mu$
    is always the chemical potential of the initial thermal state
    \cite{footnote1}.  The remaining components of retarded and
  lesser Green functions are then solved for by using separate Fourier
  transform with respect to $t$ and $t'$ in each region where both $t$
  and $t'$ do not change sign.  For the most important sector with
  both time arguments after the quench, we obtain $G^<_{++}(t,t') $
  $=$ $ G^<(t,t')\Theta(t)\Theta(t')$ by double Fourier transform,%
  \begin{subequations}\label{eq:G<pp_all}%
    \begin{align}%
      \!\!\!\!
      \tilde G^<_{++}(z,\eta)
      &=
      \int dt\,e^{izt} \int dt'\,e^{i\eta t'}G^<_{++}(t,t')
      \\
      &=
      -\int\!d\omega\,
      \frac{f(\omega)}{2\pi V^2}
      \frac{\KERN(z,\omega)+\KERN(-\eta^*,\omega)^*}{z+\eta}
      \,,\label{eq:G<pp}
      \intertext{with the abbreviations}
      \KERN(z,\omega)
      &=
      [1-\SUBKERN^{A}(z,\omega)]^{-1}-
      [1-\SUBKERN^{R}(z,\omega)]^{-1}
      \,,\label{eq:KERN}
      \\
      \SUBKERN^{\lambda}(z,\omega)
      &=
      V^2[
            w_0\,
      \gwnull^{R}_{+}(z)
      \gwnull^{\lambda}_{-}(\omega)
      +
      w_1\,
      \gweins^{R}_{+}(z)
      \gweins^{\lambda}_{-}(\omega)
      ]
      \,.
    \end{align}%
  \end{subequations}%
  Note that the initial state enters (\ref{eq:G<pp}) via the Fermi
  function, $f(\omega)$ $=$ $1/(1+e^{\beta\omega})$. Similar
  expressions are derived for the other Green functions $\Gwnull^<$
  and $\Gweins^<$~\cite{Eckstein_unpublished}.
  
  Time-dependent expectation values of observables are now obtained by
  inverse Fourier transformation and numerical integration.  Below we
  discuss the double occupation $D(t)$ $=$ $-iw_1 \Gweins^<(t,t)$ and
  the momentum distribution, i.e., the occupation $n(\epsilon,t)$ of
  single-particle eigenstates $|\epsilon\rangle$.  The latter is given
  by $n(\epsilon,t)$ $=$ $-iG^<_\epsilon(t,t)$ as defined below
  (\ref{eq:lattice-dyson-sigma}).  The total density $n_c$ is
  conserved, and the internal energy $E$ $=$ $ \langle H\rangle + \mu n_c$
  jumps by $\Delta E$ $=$ $(U_+-U_-)D(0^-)$ at the quench \cite{footnote1}.

  Simplifications occur in the limit of infinite waiting time. For $t$
  $\to$ $\infty$ the partial Fourier transformation $G^<(\omega,t)$
  $=$ $\int ds\,e^{i\omega s} G^<(t+s/2,t-s/2)$ has a well-defined limit
  $g^<_\infty(\omega)$, which is determined only by the singularity at
  $z$ $=$ $-\eta$ in (\ref{eq:G<pp}). While $G^<(\omega,t)$ is complex
  in general, its long-time limit is purely imaginary,%
  \begin{subequations}\label{eq:glpinf}%
    \begin{align}%
      g^<_{\infty} (\omega)
      &=
      \int\!d\omega'\,\frac{f(\omega')}{\pi V^2}
      i\text{Re}[\KERN(\omega+i0,\omega')]
      \label{eq:ginffromKERN}
      \\
      &=
      2\pi i h(\omega) A_+(\omega)
      \,.
    \end{align}%
  \end{subequations}%
  Plugging this result back into (\ref{eq:g-from-lambda}) and
  (\ref{eq:lambda-bethe}) we find that the steady state is
  characterized by (i) a real positive function $h(\omega)$ which
  replaces the Fermi function $f(\omega)$ in the equilibrium
  expressions, and (ii) the temperature-independent spectrum for $U_+$
  as given by $A_+(\omega)$ $=$ $\text{Im}[g^A_+(\omega)]/\pi$.  In
  particular, $E(t$ $>$ $0)$ $=$ $\int\,d\omega$ $h(\omega)$
  $(\omega+\mu)$ $A_+(\omega)$, $D_\infty $ $=$ $ w_1\int\!d\omega$
  $h(\omega)$ $\text{Im}[\gweins^A_+(\omega)]/\pi$, and
  $n_\infty(\epsilon)$ $=$ $\int\!d\omega$ $h(\omega)$
  $\text{Im}[(\omega-i0-\epsilon-\Sigma^A_+(\omega))^{-1}]/\pi$.  It
  is remarkable that subsequent quenches can be accounted for by
  simply replacing the initial occupation function $f(\omega)$ with
  the steady-state occupation function $h(\omega)$ in (\ref{eq:G<pp})
  and (\ref{eq:glpinf}).

\begin{figure}[t]
    \includegraphics[width=0.84\hsize]{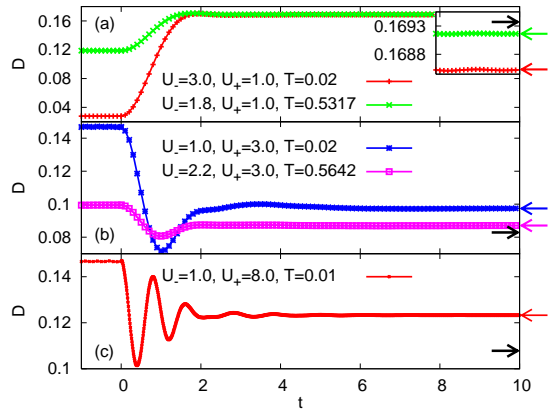}\vspace*{-3mm}
    \caption{Double occupation $D(t)$ 
      for quenches to 
      (a)~$U_+$ $=$ $1$, (b)~$U_+$ $=$ $3$, and (c)~$U_+$
      $=$ $8$, starting from an initial metallic ($U_-<2$) or
      insulating state ($U_->2$); the half-bandwidth is $2V\equiv2$.
      In (a) and (b), the internal energy
      is the same after both quenches.  Thick right-pointing arrows
      mark the double occupation in the thermal state for interaction
      $U_+$ with the same density and internal energy. These values differ from
      the stationary value $D_\infty$, marked by left-pointing arrows,
      which are approached for large times. The inset in (a) shows a
      magnification of the large-$t$ behavior.
      \label{fig:double}}
  \end{figure}

  {\em Non-thermal steady state.---} %
  In the following we focus on the case of half-filling for both $c$
  and $f$ electrons ($n_c$ $=$ $n_f$ $=$ $\frac{1}{2}$).  For these
  parameters a metal-insulator transition occurs at the critical
  interaction $U_c$ $=$ $2V$. 
  Fig.~\ref{fig:double}
  shows the double occupation $D(t)$ for different quenches, both
  within and between the two phases.  In all cases we observe
  relaxation to a new stationary value $D_\infty$ on the time
  scale~$1/V$.

  The relaxation is almost monotonic when the final interaction $U_+$
  is small (Fig.~\ref{fig:double}a), while a distinct overshoot
  (Fig.~\ref{fig:double}b) or damped oscillations
  (Fig.~\ref{fig:double}c) arise after quenches to large interactions
  ($U_+>V$).  Such transient oscillations with period $2\pi/U$ are
  expected on general grounds 
  when hopping can be neglected \cite{Greiner02b,Rigol06,Kollath07,Manmana07},  
  because the interaction part of the Hamiltonian alone leads 
  to a strictly $2\pi/U$ periodic time-evolution  
  operator $\exp(-itU\sum_i c_i^\mydagger c_i^\phdagger 
  f_i^\mydagger f_i^\phdagger)$. For small hopping $V\ll U_+$ ordinary 
  time-dependent perturbation theory then shows that the double occupancy 
  oscillates for times $t \lesssim 1/V$.

\begin{figure}[t]
    \includegraphics[width=0.84\hsize]{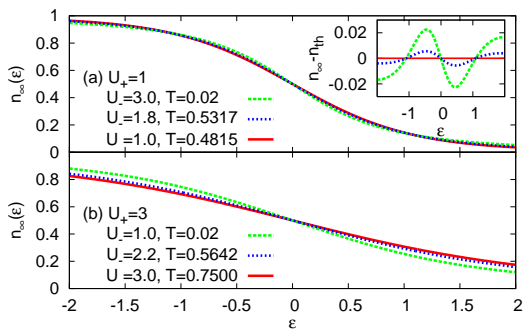}\vspace*{-3mm}
    \caption{Stationary $n_{\infty}(\epsilon)$ for quenches to
      (a)~$U_+$ $=$ $1$ and (b)~$U_+$ $=$ $3$ (same as in
      Fig.~\ref{fig:double}a and b), compared to the corresponding
      thermal values (solid red line).  
      The inset  shows a magnification of their
      differences.}
    \label{fig:nk}
  \end{figure}

  We now discuss the non-thermal character of the final steady state.
  In case of thermalization it would be fully characterized by a new
  temperature and a new chemical potential, which are fixed by density
  and internal energy only.  For Fig.~\ref{fig:double} the initial
  temperature is chosen such that the final energy $E(t$ $>$ $0)$ is
  the same for the two quenches to $U_+$ $=$ $V$
  (Fig.~\ref{fig:double}a) and also for the two quenches to $U_+$ $=$
  $3V$ (Fig.~\ref{fig:double}b).  The stationary value $D_\infty$
  clearly differs from the double occupation in the thermal state with
  the same density and 
  internal energy (thick arrows in Fig.~\ref{fig:double}a and
  b).  This lack of thermalization is also observed for the occupation
  $n(\epsilon,t)$ of single-particle states (Fig.~\ref{fig:nk}),
  for which the stationary value $n_\infty(\epsilon)$ clearly differs
  from the thermal value with the same $E$, $n_c$, and~$U_+$.
  Remarkably, thermalization does not even occur for an infinitesimal
  interaction quench $\delta U$ $=$ $U_+-U_-$ $\to$ $0$ and infinite
  waiting time. For this case we find from (\ref{eq:glpinf}) that
  $\delta g^<_\infty(\omega)$ $=$ $ -w_1
  \partial_\omega \gweins^<(\omega)\,\delta U$. For $T$ $>$ $0$ it can
  be shown \cite{Eckstein_unpublished} that $g^<_\infty(\omega)+
  \delta g^<_\infty(\omega)$ does not correspond to any thermal state
  with temperature $T+\delta T$ and chemical potential
  $\mu+\delta\mu$.


  {\em Role of constraints.---} %
  Thermalization in the FK model (\ref{eq:FK-model}) is impossible
  because the immobile $f$-particles can never find their annealed
  thermal configuration.  In addition the behavior of the
  $c$-particles is non-ergodic for any fixed configuration $\NF$ $=$
  $\{n_{f,i}\}$.  This is because for any given $\NF$ the Hamiltonian
  of the $c$ particles is quadratic, say with single-particle
  eigenstates $\ket{\alpha_\mp}$ and energies $\epsilon_{\alpha_\mp}$
  before and after the quench. As a consequence the occupation numbers
  $n_{\alpha_+}$ after the quench are time-independent and entirely
  determined by their equilibrium values before the quench,
  $n_{\alpha_+}$ $=$ $\sum_{\alpha_-} f(\epsilon_{\alpha_-})
  |\braket{\alpha_+}{\alpha_-}|^2$.

  Thermalization is prevented by this memory of the initial state that
  is frozen in $n_{\alpha_+}$.  Under this assumption the best guess
  for the steady state of the $c$ particles is a generalized Gibbs
  ensemble \cite{Rigol07}, i.e., a density matrix $\rho[\NF]$ which
  maximizes the entropy $S(\rho)=\TR(\rho\log\rho)$ subject to all the
  constraints
  given for $\expval{n_{\alpha_+}}$.  Since this $\rho[\NF]$ is a
  mixture of product states made from $\{\ket{\alpha_+}\}$, it
  predicts the site-averaged stationary Green function for a given
  configuration $\NF$ as $g^<_\infty[\NF](\omega) $ $=$ $2\pi
  \sum_{\alpha_+} \delta(\omega-\epsilon_{\alpha_+})\,n_{\alpha_+}$.
  This statistical prediction indeed agrees with the exact DMFT result
  for the infinitesimal interaction quench $\delta U$, as we now show
  using first-order perturbation theory for $|\alpha_+\rangle$.  The
  first-order energy change is $\delta\epsilon_{\alpha_-}$ $=$ $\delta
  U$ $\sum_i n_{f,i}\,\bra{\alpha_-}c_i^\mydagger
  c_i^\phdagger\ket{\alpha_-}$, while the change of $n_{\alpha_+}$ is
  of order $\delta U^2$. This gives $\delta g^<_{\infty}[\NF](\omega)$
  $=$ $-2\pi\partial_\omega$ $\sum_{\alpha_-}
  \delta(\omega-\epsilon_{\alpha_-}) f(\epsilon_{\alpha_-})
  \delta\epsilon_{\alpha_-}$ $=$ $-w_1\partial_\omega
  \gweins^<[\NF](\omega)\,\delta U$.  Because the probabilities
  $P[\NF]$ of the $f$ configurations are time-independent and depend
  only on the initial state of the $c$ electrons, averaging over $\NF$
  recovers our DMFT result for $\delta g^<_{\infty}(\omega)$.
  Thus generalized Gibbs ensembles provide the appropriate
  statistical description of this final steady state, at 
  least for simple observables.  In this aspect
  our results, which are strictly valid in infinite dimensions,
  resemble those for one-dimensional integrable models
  \cite{Rigol07,Rigol06,Cazalilla06}. 
  

  {\em Conclusion.---} %
  The exact DMFT solution of the FK model after an interaction quench
  shows that this isolated many-body system relaxes to a new steady
  state. The momentum occupation and double occupation in the final
  state do not correspond to any thermal state.  Instead these
  observables are described by means of generalized Gibbs ensembles,
  averaged over all $f$ configurations.
  
  In general, DMFT has been very successful for correlated systems in
  equilibrium and gives a good description of local observables in
  three-dimensional systems.  Its application to non-equilibrium
  phenomena is thus very promising, and DMFT results for quenches in
  the Hubbard model would be desirable. If the Hubbard model indeed
  thermalizes, as expected for a non-integrable
  system~\cite{Rigol07b}, this would lead to a crossover between
  ergodic and non-ergodic regimes. This crossover could be studied
  experimentally with ultracold atomic gases in optical lattices,
  e.g., with mixtures of polarized fermionic atoms 
  for which the lattice depth can be tuned separately.


  We thank D.~Vollhardt, K.~Byczuk, and M.~Rigol for
  useful discussions.  M.E.\ acknowledges support by Studienstiftung
  des Deutschen Volkes.  This work was supported in part by the SFB
  484 of the DFG.



\begin{thebibliography}{99}
\bibitem{footnote1}%
    Note that adding a time-dependent single-particle potential term
    $\epsilon(t) N$ at $t$ $=$ $0$ leads to phase factors
    $\exp[i\int_{t}^{t'}d\tau$ $\epsilon(\tau)]$ in the Green
    functions, and thus is of no consequence for equal-time $(t$
      $=$ $t')$ observables.
\end{thebibliography}
\end{document}